\documentclass[aps,12pt,preprint]{revtex4}
\usepackage{amsmath}
\usepackage{bm}

\begin{document}
\preprint{BA-05-06}

\title{The Origin of a Peculiar Extra $U(1)$}
\author{S.M. Barr}
\email{smbarr@bxclu.bartol.udel.edu}
\affiliation{Bartol Research Institute\\University of Delaware\\
Newark, DE 19716}
\author{I. Dorsner}
\email{idorsner@ictp.trieste.it} \affiliation{The Abdus Salam
International Centre for Theoretical Physics\\
Strada Costiera 11, 31014 Trieste, Italy}
\begin{abstract}

The origin of a family-independent ``extra $U(1)$", discovered by
Barr, Bednarz, and Benesh and independently by Ma, and whose
phenomenology has recently been studied by Ma and Roy, is
discussed. Even though it satisfies anomaly constraints in a
highly economical way, with just a single extra triplet of leptons
per family, this extra $U(1)$ cannot come from four-dimensional
grand unification. However, it is shown here that it can come from
a Pati-Salam scheme with an extra $U(1)$, which explains the
otherwise surprising cancellation of anomalies.
\end{abstract}
 \maketitle
The possibility of ``extra U(1)" gauge groups has been much
discussed. By extra $U(1)$ is meant a $U(1)$ factor in addition to
the Standard Model group $G_{SM} = SU(3)_c \times SU(2)_L \times
U(1)_Y$. If the extra $U(1)$ (often called $U(1)'$) is broken near
the weak scale, then the ``extra $Z$" gauge boson associated with
it (often called $Z'$) can lead to interesting phenomenology
\cite{extrazphen,
Robinett:1981yz,Leung:1984qa,Durkin:1985ev,Barger:1985dd,
London:1986dk,Amaldi:1987fu,Costa:1987qp,Hewett:1988xc,
Mahanthappa:1991pw,Langacker:1991pg,adh,cddt}.

Grand unified theories (GUTs)  based on groups with rank greater
than 4 can lead to such $U(1)$ factors. For instance, $SO(10)$
contains an extra $U(1)$ that could survive to low energies
(compared to the GUT scale), and $E_6$ contains two extra $U(1)$
symmetries, both of which, or some linear combination of which,
could survive. These particular $U(1)'$ groups and their
associated $Z'$ bosons have been especially well studied, and in
fact are the standard examples of extra $U(1)$ groups. In this
paper we discuss a peculiar $U(1)'$ that is quite simple and
economical, but does not come from grand unification. The
existence of this $U(1)'$ was first discovered in a little-known
paper by one of the present authors \cite{anom}, (see Eq.\ (30)
and the conclusions of that paper). Recently, it has been
rediscovered by E.\ Ma \cite{ma}, and its phenomenology studied by
E.\ Ma and D.P.\ Roy \cite{maroy}. Here we explain how this
$U(1)'$ can arise in a simple Pati-Salam scheme of quark-lepton
unification, which explains group-theoretically speaking, why this
surprising anomaly-free $U(1)'$ exists. We also discuss how it may
be related to grand unified models in higher space-time
dimensions.

The existence of a $U(1)'$ brings with it additional anomaly
cancellation conditions. (These are given a very general analysis
in \cite{anom}.) If the group is $G_{SM} \times U(1)'$ there are
six new anomaly conditions, which in an obvious notation are the
$(3_c)^2 1'$, $(2_L)^2 1'$, $(1_Y)^2 1'$, $1_Y (1')^2$, $(1')^3$,
and $(gravity) 1'$ conditions. (For a general analysis of
solutions to these conditions, see \cite{anom}.) If $U(1)'$ is
family-independent and there are $N$ multiplets of $G_{SM}$ per
family, then there are $N$ quantities (i.e.\ the $U(1)'$ charges
of these multiplets) that have to satisfy six conditions. At first
glance, it would seem that there would be solutions if $N \geq 6$.
However, in fact $N$ must be at least $8$, generally speaking, for
a solution to exist. The reason is the following. First, the
anomaly conditions are homogeneous and therefore do not fix the
overall normalization of the $U(1)'$ charges of the fermion
multiplets. Second, given any non-trivial solution for the $U(1)'$
charges, say $Q'(f) = X_f$ (where $Q'$ is the generator of $U(1)'$
and $f$ is any fermion), then $Q'(f) = X_f + \alpha Y(f)$ must
also be a solution, where $Y$ is the hypercharge and $\alpha$ is a
free parameter. (Note that the $U(1)'$ charges given for the
extra-lepton-triplet $U(1)$ in \cite{anom} are in fact a linear
combination of the charges given here with hypercharge.) Since two
continuous parameters do not get fixed, there must be 8 rather
than six charges, generally speaking, for a non-trivial solution
to the anomaly conditions to exist, and thus 8 multiplets per
family. (No matter how many multiplets there are, there is always
the trivial solution $Q'(f) = \alpha Y(f)$~\cite{anom,seq,seq1}.)

Let us now ask what happens if there is only {\it one\/} extra
fermion multiplet per family in addition to the usual five (namely
$(u,d)_L$, $u^c_L$, $d^c_L$, $(\nu, e^-)_L$, and $e^+_L$). From
the foregoing, it follows that there should not be any non-trivial
solutions for a $U(1)'$ except, perhaps, in special cases. In
fact, there are exactly two special cases. One is the well-known
extra $U(1)$ that comes from $SO(10)$. The other is the peculiar
$U(1)$ that we are discussing in this paper.

If there is only one extra fermion multiplet per family, then
(assuming the gauge quantum numbers of all the families to be the
same) the extra multiplet must be a in a real representation of
$G_{SM}$. Call it $(R_3, R_2, 0)$. Define the ratios $r_3 \equiv
C_3(R_3)/R_3$ and $r_2 \equiv C_2 (R_2)/R_2$ (where the Casimirs
are normalized so that $C_3(3) = C_2(2) = \frac{1}{2}$). It can be
shown that there is no solution to the complete set of anomaly
equations unless two non-trivial conditions are satisfied by the
dimensions and Casimirs of the extra multiplet:

\begin{equation}
\label{1} 0 = - \frac{44}{9} r_3^2 + 2 r_3 r_2 + \frac{3}{2} r_2^2
+ \frac{4}{3} r_3 - r_2,
\end{equation}

\noindent and

\begin{equation}
\label{2} 0 = - \frac{32}{3} r_3^3 + \frac{61}{9} r_3^2 r_2 +
\frac{13}{2} r_3 r_2^2 + r_2^3 + 4 r_3^2 - \frac{17}{3} r_3 r_2 -
2 r_2^2 + r_2 + \frac{1}{12} \left( \frac{1}{R_3^2 R_2^2} - 1
\right).
\end{equation}

\noindent (These equations can be derived by solving the four
linear anomaly conditions to express all the $U(1)'$ charges in
terms of two parameters, and then substituting the result into the
quadratic and cubic anomaly conditions. When this is done, the
unfixed parameters drop out, as they should, and Eqs.~\eqref{1}
and \eqref{2} result.) That two non-trivial conditions have to be
satisfied by the choice of representation just follows from the
fact that there are only 6 multiplets rather than the 8 per family
that is needed in the general case for a solution.

If the extra multiplet is a color singlet, then $R_3 = 1$, $r_3
=0$, and the equations reduce to $(\frac{3}{2} r_2 - 1) r_2 = 0$
and $r_2^3 - 2 r_2^2 + r_2 + \frac{1}{12}(R_2^{-2} - 1) = 0$.
These have only two solutions: either $R_2 = 1$ ($\Rightarrow r_2
= 0$), or $R_2 = 3$ ($\Rightarrow r_2 = \frac{2}{3}$). There are
no solutions if $R_3$ is a non-singlet. Thus, the only two
solutions are the following \cite{ma}:

\vspace{0.2cm}

\noindent {\bf Solution 1:} If the extra fermion multiplet in each
family is just a $(1,1,0)$ of $G_{SM}$ (i.e.\ a ``right-handed
neutrino"), then there is the solution $Q'((u,d)_L$, $u^c_L,
d^c_L, (\nu, e^-)_L, e^+_L, N^c_L) = (+1, +1, -3, -3, +1, +5)$.
This solution can be understood group-theoretically: the $Q'$ is
just the generator of the extra $U(1)$ contained in $SO(10)$.

\vspace{0.2cm}

\noindent {\bf Solution 2:} If the extra multiplet in each family
is a $(1,3,0)$ of $G_{SM}$, i.e. a lepton triplet, which we shall
denote $(t^+, t^0, t^-)$, then there is the following simple
solution which we will call the ``extra-lepton-triplet $U(1)$" or
$U(1)_{ELT}$:

\begin{equation}
\label{3}
\begin{array}{cccc}
Q' = +1: & \left( \begin{array}{c} u \\ d \end{array} \right)_L, &
\left( \begin{array}{c} \nu \\ e^- \end{array} \right)_L,
& e^+_L, \\ & & & \\
Q' = -1: & u^c_L, & d^c_L, & \left(
\begin{array}{c} t^+ \\ t^0 \\ t^- \end{array} \right)_L.
\end{array}
\end{equation}

\noindent The $(3_c)^2 1'$ anomaly condition is obviously
satisfied, since the contributions from the $u$ and $d$ cancel
those from the $u^c$ and $d^c$. The $(1')^3$ and $(gravity) 1'$
conditions are both satisfied because there happen to be nine (=
6+2+1) fields with $Q' = +1$ and nine (= 3 + 3 + 3) fields with
$Q' = -1$. The $(2_L)^2 1'$ condition is satisfied because the
contribution of the $SU(2)_L$ triplet $(t^+, t^0, t^-)_L$ cancels
the contribution of the four (counting color) $SU(2)_L$ doublets
$(u,d)_L$ and $(\nu, e^-)_L$ (since $C_2(3) = 4 C_2(2)$). Finally,
the $(1_Y)^2 1'$ condition is satisfied because $6 (\frac{1}{6})^2
+ 2 (- \frac{1}{2})^2 + 1( +1)^2 - 3 (- \frac{2}{3})^2 - 3 (
\frac{1}{3})^2 - 3 (0)^2 = \frac{5}{3} - \frac{5}{3} = 0$.

As one can see, these anomaly cancellations are non-trivial, and
it seems rather remarkable that such a solution exists, since this
set of fermions and charges is {\it not\/} embeddable in a grand
unified theory, unlike the previous case. (It is not difficult to
see why it is not embeddable in a grand unified theory. An
$SU(2)_L$ triplet is in the symmetric product of two doublets, and
thus can come from a multiplet that is in a symmetric product of
two ${\bf 5}$s of $SU(5)$. However, such an $SU(5)$ multiplet will
also contain a ${\bf 6}$ of color.) It thus appears at first
glance that the existence of this solution to the anomaly
conditions is merely a fluke. However, as we shall now see, this
solution can be related to a Pati-Salam model \cite{ps} with an
extra $U(1)$ in a way that makes it seem a little less surprising.

Suppose we consider a model with the group $G_{PS} \times U(1)' =
SU(4)_c \times SU(2)_L \times SU(2)_R \times U(1)'$ and fermion
multiplets

\begin{equation}
\label{4}
\begin{array}{cll}
Q' = +1: & \left( \begin{array}{cc} u & \nu \\ d & e^- \end{array}
\right)_L = (4,2,1)^{+1}_L, & \left( \begin{array}{c} e^+ \\ T^0
\\ T^-
\end{array} \right)_L = (1,1,3)^{+1}_L, \\ & & \\
Q' = -1: & \left( \begin{array}{cc} u^c & T^{c0} \\ d^c & T^+
\end{array}
\right)_L = (\overline{4},1,2)^{-1}_L, & \left( \begin{array}{c}
t^+ \\ t^0 \\ t^- \end{array} \right)_L = (1,3,1)^{-1}_L.
\end{array}
\end{equation}

\noindent Note that the Pati-Salam multiplets $(4,2,1) +
(\overline{4},1,2)$ {\it cannot\/} come from a spinor of $SO(10)$
because they have opposite $U(1)'$ charges. Thus this model cannot
come from a grand unified model (though, as we shall see, it can
be related to one in higher dimensions). The  $(4_c)^2 1'$,
$(1')^3$, and $(gravity) 1'$ anomalies cancel in an obvious way.
The $(2_L)^2 1'$ and $(2_R)^2 1'$ anomalies cancel because (as
noted above) one $SU(2)$ triplet contributes with the same weight
as four doublets. There are in this model five anomaly conditions
that involve $U(1)'$, and they are satisfied by only three charge
ratios. Nevertheless, the anomalies cancellation here is more
transparent than in Eq.~\eqref{3}.

Now imagine that a Higgs field $\Omega$ in a $(4,1,2)^0$
representation of $G_{PS} \times U(1)'$ has a Yukawa coupling to
the fermion multiplets $(\overline{4}, 1,2)^{-1}_L$ and
$(1,1,3)^{+1}_L$. If its electrically neutral component gets a
vacuum expectation value of magnitude $M_R$, then the pairs
$(T^{c0} T^0)$ and $(T^+ T^-)$ will obtain mass of order $M_R$,
and the group $G_{PS} \times U(1)'$ will break to $G_{SM} \times
U(1)'$ with exactly the residual light fermion content shown in
Eq.~\eqref{3}.

In the $SU(3)_c \times SU(2)_L \times U(1)_Y \times U(1)'$ model,
lepton masses arise from the following matrices:

\begin{equation}
\label{5}
\begin{array}{ccl}
{\cal L}_{lepton} & = & \left( e^+_i, t^+_i \right) \left(
\begin{array}{cc} \gamma_{ij} \langle H_2^{(-2)} \rangle &
\delta_{ij} \langle H_3^{(0)} \rangle \\ \sqrt{2} \beta_{ij}
\langle H_2^{(0)} \rangle & \alpha_{ij} \langle H_1^{(2)} \rangle
\end{array} \right) \left( \begin{array}{c} e^-_j \\ t^-_j \end{array}
\right) \\ & & \\
& + & \frac{1}{2} \left( \nu , t^0 \right) \left(
\begin{array}{cc} \kappa_{ij} \langle H_3^{(-2)} \rangle &
\beta^T_{ij} \langle H_2^{(0)} \rangle \\ \beta_{ij} \langle
H_2^{(0)} \rangle & \alpha_{ij} \langle H_1^{(2)} \rangle
\end{array} \right) \left( \begin{array}{c} \nu \\ t^0 \end{array}
\right).
\end{array}
\end{equation}

\noindent The $i,j$ are family indices. The notation $H_{R}^{(m)}$
refers to a Higgs field in an $R$-component multiplet of $SU(2)_L$
(i.e.\ one with weak isospin $\frac{R-1}{2}$) that has $U(1)'$
charge $Q' = m$. The $\rho$ parameter tells us that the VEVs of
$SU(2)_L$ triplet Higgs, if there are any, must be small compared
to the weak scale. Let us therefore neglect the entries
proportional to $H_3^{(0)}$ and $H_3^{(-2)}$ in Eq.~\eqref{5}. The
singlet VEV $\langle H_1^{(2)} \rangle$ must, on the other hand,
be large compared to the weak scale, since it generates the masses
of the triplet leptons and the mass of the $Z'$. How large depends
on the entries in Eq.~\eqref{5} proportional to $\langle H_2^{(0)}
\rangle$. There are two interesting cases.

\vspace{0.2cm}

\noindent {\bf Case 1.} If the entries proportional to $\langle
H_2^{(0)} \rangle$ in Eq.~\eqref{5} are of order the weak scale,
then it must be that $\langle H_1^{(2)} \rangle$ is superlarge (of
order $10^{14}$\,GeV or more) in order to keep the observed
neutrino masses small. This gives the usual neutrino seesaw
mechanism, except that the superheavy ``right-handed neutrinos"
are part of triplets of superheavy leptons. In this case, since
$U(1)'$ is broken at superlarge scales, there is no interesting
$Z'$ phenomenology, and the mixing of the known leptons with the
triplet leptons is negligible. There could, however, be
interesting consequences for leptogenesis.

\vspace{0.2cm}

\noindent {\bf Case 2.} If the entries proportional to $\langle
H_2^{(0)} \rangle$ in Eq.~\eqref{5} vanish (or are extremely small
compared to the weak scale), then $\langle H_1^{(2)} \rangle$ can
have any value from the Planck scale down to about a TeV. If the
scale of $U(1)'$ breaking is in the TeV range, there would be
interesting $Z'$ phenomenology, which has been analyzed very
thoroughly in \cite{ma}, \cite{maroy}. It should be observed that
a $U(1)'$-neutral Higgs doublet $H_2^{(0)}$ is needed to give mass
to the up quarks and down quarks, as can be seen from
Eq.~\eqref{1}. Consequently, if the entries in Eq.~\eqref{5} that
are proportional to $\langle H_2^{(0)} \rangle$ are to vanish or
be small {\it naturally}, some symmetry reason for this
suppression must exist. One such reason could be a $Z_2$ symmetry
under which the triplet leptons are odd and all other quarks and
leptons (and all Higgs fields) are even. Note that this would
prevent the triplet leptons from mixing with the ordinary leptons,
and render them stable. They might then be candidates for dark
matter. It is also possible, however, to construct models in which
there is significant mixing of the triplet leptons and ordinary
doublet and singlet leptons, for example through small entries
proportional to $\langle H_3^{(0)} \rangle$ or through higher
dimension operators. This mixing could lead to observable
violations of the universality of the weak interactions of the
leptons.

Obviously the ideas mentioned here (some of which were discussed
in great detail in \cite{ma}, \cite{maroy}) do not exhaust all the
possibilities, but they illustrate some of the ways in which
lepton phenomenology could be different if there is the extra
lepton triplet $U(1)$.

Another way in which these lepton triplets could be significant is
in their effect on the running of the gauge couplings. In fact,
they could allow unification of the Standard Model gauge couplings
without low-energy supersymmetry. This may seem inconsistent with
the fact, noted earlier, that the $U(1)_{ELT}$ cannot come from
grand unification. However, it is consistent with
non-supersymmetric unification of the standard model gauge groups
in extra dimensions~\cite{Kawamura:1999nj,Kawamura:2000ir}. For
example, suppose a fifth dimension is compactified on a $S_1/Z_2$
orbifold, in such a way that an $SO(10) \times U(1)'$ gauge
symmetry in the bulk is broken by the compactification to the
$G_{PS} \times U(1)'$ subgroup. This happens if the gauge bosons
of the coset $SO(10)/G_{PS}$ are odd under the $Z_2$. On the
branes that bound the bulk, since they are fixed points of the
$Z_2$, there is only the gauge symmetry $G_{PS} \times U(1)'$.
Thus, the quarks and leptons living on one of these branes can
have any $G_{PS} \times U(1)'$ quantum numbers consistent with
anomaly cancellation. In particular, they could have the quantum
numbers shown in Eq.~\eqref{4}. The Standard Model gauge couplings
(to the extent that one can neglect corrections from gauge kinetic
terms on the branes) would be unified due the $SO(10)$ symmetry in
the bulk. It should be noted that even if the gauge unification
happened at a scale much below $10^{15}$\,GeV proton decay would
not be a problem, since the gauge-mediated proton decay comes from
gauge bosons in the coset $SO(10)/G_{PS}$, which do not live on
the branes where the quarks and leptons exist

In such a brane scenario, it is not necessary that all three
families have the structure shown in Eq.~\eqref{4}. It could be,
for instance, that some of the families contain lepton triplets
and the $U(1)'$ charges of Eq.~\eqref{4}, while the remaining
families have no lepton triplets and are neutral under $U(1)'$.
These latter families could exist either on the branes or in the
bulk. However, if they existed in the bulk or mixed with fermions
in the bulk, one would have to worry about the proton decay rate.

Assuming unification of gauge couplings fixes the mass of the
triplet leptons (assuming all the triplet leptons have nearly the
same mass). The mass of the triplets depends, of course, on how
many of the families have triplets. In numerical examples
presented in Table~\ref{tab:table1} we consider running of the
gauge couplings at the one-loop level between the electroweak
scale and grandunifying scale with $n_3$ extra triplets. We impose
exact unification and use central values for $\sin^2 \theta_w$,
$\alpha_{em}^{-1}$ and $\alpha_{s}$ as given in~\cite{PDG2004} to
determine intermediate scale where triplets reside. Clearly,
similar analysis can also be performed in the orbifold scenario if
the full particle content of the bulk and brane(s) is specified.

\begin{table}[h]
\caption{\label{tab:table1} Triplet mass versus number of triplets
($n_3$) in case of one ($n=1$) and two ($n=2$) light Higgs
doublets. Grandunifying scale is $2.5 \times 10^{14}$\,GeV and
$1.9 \times 10^{14}$\,GeV respectively. }
\begin{ruledtabular}
\begin{tabular}{ccc}
   \hline
     & $n=1$ & $n=2$ \\
   \hline
   $n_{3}=1$ & $7.2 \times 10^{6}$\,GeV & $9.3 \times 10^{7}$\,GeV \\
   $n_{3}=2$ & $4.2 \times 10^{10}$\,GeV & $1.3 \times 10^{11}$\,GeV \\
   $n_{3}=3$ & $7.6 \times 10^{11}$\,GeV & $1.5 \times 10^{12}$\,GeV \\
   \hline
\end{tabular}
\end{ruledtabular}
\end{table}
\vspace{0.2cm}

Of course, it is possible that there could be extra lepton
triplets at low or intermediate energy without there being any
extra $U(1)'$ gauge group. However, if the only symmetry is
$G_{SM}$, then there is nothing to ``protect" the extra triplets
from acquiring superlarge mass, since they are in a real
representation of $G_{SM}$. The extra lepton triplet $U(1)$ does
protect these triplets from acquiring such masses. Of course, in a
non-supersymmetric model, the mass of the Higgs doublet is
fine-tuned anyway; so one could argue that nothing is lost by
fine-tuning the masses of the lepton triplets to be small also.
However, the Higgs mass fine-tuning could be explained, in
principle, ``anthropically", whereas it is less obvious that the
low masses of lepton triplets could be explained so easily in that
way.

In conclusion, we have shown that the remarkable extra $U(1)$ that
was discovered in \cite{anom} and \cite{ma} and analyzed in
\cite{maroy} has a simple group-theoretical explanation in terms
of Pati-Salam unification, and may even be related to grand
unification in higher dimensions. If it is related to grand
unification in higher dimensions, then the scale at which the
$U(1)'$ is broken is too large for the $Z'$ phenomenology to be
directly accessible in the near future. However, the extra heavy
leptons would still be relevant to neutrino physics and possibly
to leptogenesis. If the scheme comes from Pati-Salam unification,
but not grand unification, then the scale of the $U(1)'$ breaking
can have any value, and can, in particular, be near the TeV scale.



\begin{thebibliography}{99}

\bibitem{extrazphen}
   N.~G.~Deshpande and D.~Iskandar,
   Phys.\ Rev.\ Lett.\  {\bf 42}, 20 (1979).

\bibitem{Robinett:1981yz}
   R.~W.~Robinett and J.~L.~Rosner,
   Phys.\ Rev.\ D {\bf 25}, 3036 (1982)
   [Erratum-ibid.\ D {\bf 27}, 679 (1983)].

\bibitem{Leung:1984qa}
   C.~N.~Leung and J.~L.~Rosner,
   Phys.\ Rev.\ D {\bf 29}, 2132 (1984).

\bibitem{Durkin:1985ev}
   L.~S.~Durkin and P.~Langacker,
   Phys.\ Lett.\ B {\bf 166}, 436 (1986).

\bibitem{Barger:1985dd}
   V.~D.~Barger, N.~G.~Deshpande and K.~Whisnant,
   Phys.\ Rev.\ Lett.\  {\bf 56}, 30 (1986).

\bibitem{London:1986dk}
   D.~London and J.~L.~Rosner,
   Phys.\ Rev.\ D {\bf 34}, 1530 (1986).

\bibitem{Amaldi:1987fu}
   U.~Amaldi {\it et al.},
   Phys.\ Rev.\ D {\bf 36}, 1385 (1987).

\bibitem{Costa:1987qp}
   G.~Costa, J.~R.~Ellis, G.~L.~Fogli, D.~V.~Nanopoulos and F.~Zwirner,
   Nucl.\ Phys.\ B {\bf 297}, 244 (1988).

\bibitem{Hewett:1988xc}
   J.~L.~Hewett and T.~G.~Rizzo,
   Phys.\ Rept.\  {\bf 183}, 193 (1989).

\bibitem{Mahanthappa:1991pw}
   K.~T.~Mahanthappa and P.~K.~Mohapatra,
   Phys.\ Rev.\ D {\bf 43}, 3093 (1991)
   [Erratum-ibid.\ D {\bf 44}, 1616 (1991)].

\bibitem{Langacker:1991pg}
   P.~Langacker and M.~x.~Luo,
   Phys.\ Rev.\ D {\bf 45}, 278 (1992).

\bibitem{adh}
   T.~Appelquist, B.~A.~Dobrescu, and A.~R.~Hopper,
   Phys.\ Rev.\ D {\bf 68}, 035012 (2003).

\bibitem{cddt}
   M.~Carena, A.~Daleo, B.~A.~Dobrescu, and T.~M.~P.~Tait,
   Phys.\ Rev.\ D {\bf 70}, 093009 (2004).


\bibitem{anom}
   S.~M.~Barr, B.~Bednarz and C.~Benesh,
   Phys.\ Rev.\ D {\bf 34}, 235 (1986).

\bibitem{ma}
    E.~Ma,
    Mod.\ Phys.\ Lett.\ A {\bf 17}, 535 (2002).

\bibitem{maroy}
    E.~Ma and D.~P.~Roy,
    Nucl.\ Phys.\ B {\bf 644}, 290 (2002).

\bibitem{seq}
   K.~T.~Mahanthappa and P.~K.~Mohapatra,
   Phys.\ Rev.\ D {\bf 42}, 2400 (1990).

\bibitem{seq1}
   V.~A.~Kostelecky and S.~Samuel,
   Phys.\ Lett.\ B {\bf 270}, 21 (1991).

\bibitem{ps}
   J.~C.~Pati and A.~Salam,
   Phys.\ Rev.\ D {\bf 10}, 275 (1974).

\bibitem{Kawamura:1999nj}
   Y.~Kawamura,
   Prog.\ Theor.\ Phys.\  {\bf 103}, 613 (2000).

\bibitem{Kawamura:2000ir}
   Y.~Kawamura,
   Prog.\ Theor.\ Phys.\  {\bf 105}, 691 (2001).

\bibitem{PDG2004}
   S.~Eidelman {\it et al.},
   Phys.\ Lett.\ B {\bf 592}, 1 (2004).

\end{thebibliography}
\end{document}